\documentclass{revtex4}

\usepackage{graphicx}

\begin{document}

\begin{abstract}
F$_\mathrm{o}$F$_1$-ATP synthase is a factory for synthesizing ATP in virtually all cells. 
Its core machinery is the subcomplex F$_1$-motor (F$_1$-ATPase) and performs the reversible mechanochemical coupling. 
Isolated F$_1$-motor hydrolyzes ATP, which is accompanied by unidirectional rotation of its central $\gamma$-shaft. 
When a strong opposing torque is imposed, the $\gamma$-shaft rotates in the opposite direction and drives the F$_1$-motor to synthesize ATP. 
This mechanical-to-chemical free-energy transduction is the final and central step of the multistep cellular ATP-synthetic pathway. 
Here, we determined the amount of mechanical work exploited by the F$_1$-motor to synthesize an ATP molecule during forced rotations using methodology combining a nonequilibrium theory and single molecule measurements of responses to external torque. 
We found that the internal dissipation of the motor is negligible even during rotations far from a quasistatic process. 
\end{abstract}

\title{Single molecule thermodynamics of ATP synthesis by F$_1$-ATPase}

\author{Shoichi Toyabe}
\affiliation{Faculty of Science and Engineering, Chuo University, Tokyo 112-8551, Japan}
\affiliation{Graduate School of Engineering, Tohoku University, Sendai 980-8579, Japan}
\author{Eiro Muneyuki}
\email{emuneyuk@phys.chuo-u.ac.jp}
\affiliation{Faculty of Science and Engineering, Chuo University, Tokyo 112-8551, Japan}

\maketitle

\setlength{\baselineskip}{0.7cm}

\vspace{0.5cm}

F$_\mathrm{o}$F$_1$-ATP synthase comprises an F$_\mathrm{o}$-motor embedded in a membrane (inner membrane of the mitochondria in the eukaryotic cells) and an F$_1$-motor protruding from the membrane (Fig. \ref{fig:Intro}a). Proton translocation through the F$_\mathrm{o}$-motor driven by the transmembrane electrochemical potential unidirectionally rotates the c-ring of the F$_\mathrm{o}$-motor. 
Because the c-ring is connected to the $\gamma$-shaft of F$_1$-motor, the c-ring imposes torque on the $\gamma$-shaft causing it to rotate. 
The forced rotation of the $\gamma$-shaft induces the F$_1$-motor's stator $\alpha_3\beta_3$ ring to synthesize ATP from ADP and P$_\mathrm{i}$ (inorganic phosphate). 
The F$_1$-motor converts the mechanical work transferred from the c-ring to the chemical free energy of ATP synthesis from ADP and P$_\mathrm{i}$, $\Delta\mu$ \cite{Itoh2004, Rondelez2005}. 
Approximately 95\% of cellular ATP is synthesized by this rotary mechanochemical transduction. 
In contrast, the isolated F$_1$-motor hydrolyzes ATP to ADP and phosphate and rotates the $\gamma$-shaft by converting $\Delta\mu$ to mechanical motion \cite{Noji1997,Yasuda1998}. 
However, the rotational direction is opposite to that of the ATP-synthetic rotation. 
The $\gamma$-shaft rotates 120$^\circ$ per ATP hydrolysis \cite{Rondelez2005, Yasuda1998}.
Thus, the F$_1$-motor reversibly transduces free-energy between mechanical work and $\Delta\mu$ \cite{Abrahams1994, Boyer1993,Noji1997,Yasuda1998}.

The detailed kinetics and reaction scheme of the F$_1$-motor's rotations have been studied intensively. 
In contrast, knowledge regarding the energetics is limited \cite{Muneyuki2007Biophys, Toyabe2010PRL, Toyabe2011, Toyabe2012, Toyabe2013, Yasuda1998} due to the difficulty in studying the energetics of such nanosized engines working at an energy scale comparable to thermal energy. 
An experimental methodology was recently established by combining measurements of single molecule responses with nonequilibrium theory \cite{Toyabe2013, Toyabe2012, Toyabe2010PRL, Toyabe2011}.
Previous research revealed the high efficiency of the ATP-hydrolytic rotations by the isolated F$_1$-motor as follows: (i) The maximum work performed by the F$_1$-motor during a 120$^\circ$ rotation is equal to $\Delta\mu$ within experimental error \cite{Toyabe2011}. Specifically, the F$_1$-motor achieves a unity thermodynamic efficiency at the stalled state where the mean rotational rate vanishes because of a hindering external torque. 
(ii) The F$_1$-motor converts $\Delta\mu$ to mechanical work with negligible internal heat dissipation even during rotations far from quasistatic process \cite{Toyabe2010PRL}. 
(iii) The motor minimizes the internal heat dissipation by switching its chemical state depending on the angular position of the $\gamma$-shaft \cite{Toyabe2012}.

However, the primary physiological role played by the F$_1$-motor is not hydrolysis but mechanochemical ATP synthesis driven by the forced-rotations of the $\gamma$-shaft. 
This raises the fundamental unanswered question regarding the free-energy transduction as follows: How much work is required to drive the F$_1$-motor to synthesize an ATP molecule? 
To address this question, we improved available methodology \cite{Toyabe2010PRL} and investigated the energetics of the forced ATP-synthetic rotations by the isolated F$_1$-motor.

\section*{Experiment}

\begin{figure}[!h]
\centerline{\includegraphics[scale=1]{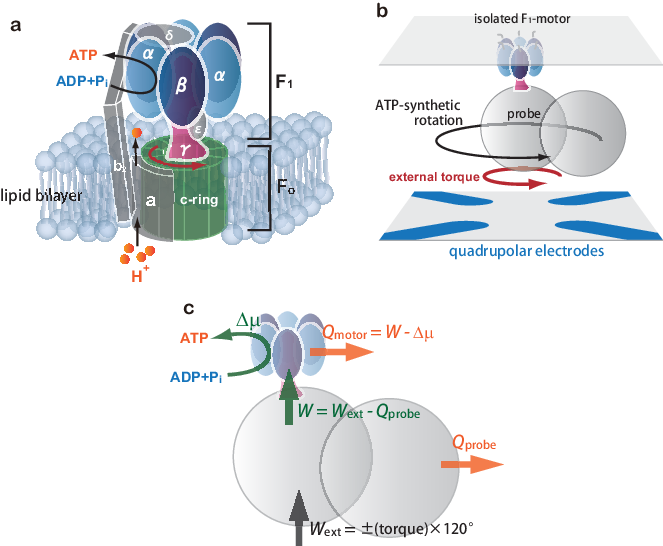}}
\caption{\label{fig:Intro}
The structure of F$_\mathrm{o}$F$_1$-ATP synthase and a model for catalysis. 
{\bf a}, F$_\mathrm{o}$F$_1$-ATP synthase. Proton flux through the F$_\mathrm{o}$-motor rotates the c-ring and the $\gamma$-shaft of the F$_1$-motor and drives ATP synthesis. 
{\bf b}, Single molecule response measurement of the isolated $\alpha_3\beta_3\gamma$ subcomplex of the F$_1$-motor as the model experiment system of the F$_\mathrm{o}$F$_1$-ATP synthase. 
The F$_1$-motor was fixed on the upper cover slip. 
A dimeric polystyrene particle (diameter = 300 nm) was attached to the $\gamma$-shaft with an elastic protein linker (streptavidin). 
Torque was imposed on the particle by an electrorotation method using quadrupolar electrodes patterned on the lower glass slide. A strong hindering torque rotates the $\gamma$-shaft in the opposite direction and drives the F$_1$-motor to synthesize ATP. 
{\bf c}, Energy flow through the F$_1$-motor.
}
\end{figure}

An isolated F$_1$-motor molecule was adhered to an upper glass surface (Fig. \ref{fig:Intro}b) \cite{Noji1997, Yasuda1998}. 
A submicron particle was attached to the $\gamma$-shaft with an elastic protein linker to probe the $\gamma$-shaft rotations by an optical microscope. 
We imposed a torque on the probe using an electrorotation method \cite{Berry1995,Toyabe2010PRL, Toyabe2011, Washizu1991, Watanabe-Nakayama2008}.
The electrorotation method uses a high-frequency alternating-current electric field generated by quadrupolar electrodes and imposes a torque with a controlled magnitude (see Methods). 
Thus, we can measure the rotational response of a single motor molecule as a function of external torque. 
This system mimics the mechanochemical transduction inside the F$_\mathrm{o}$F$_1$-ATP synthase; the upper glass surface, the probe, and the torque imposed correspond to the stator, the c-ring, and the proton-motive force that drives the c-ring, respectively.

Consider the energetics of this system as follows (Fig. \ref{fig:Intro}c): Under sufficiently strong torque in the ATP-synthetic direction, the probe-shaft complex rotates in that direction \cite{Toyabe2011}. 
The amount of the external torque multiplied by 120$^\circ$ with a sign depending on the rotational direction is the work, $W_\mathrm{ext}$, performed on the probe per ATP cycle. 
A portion of $W_\mathrm{ext}$, $W$, is transferred to the motor by rotating the $\gamma$-shaft with the elastic linker and the remainder, $Q_\mathrm{probe}$, dissipates through the viscous friction of the probe.
\begin{equation}\label{eq:balance}
W=W_\mathrm{ext}-Q_\mathrm{probe}.   
\end{equation}
The motor receives $W$, converts it to the chemical free energy change $\Delta\mu$ by synthesizing an ATP molecule at its $\alpha_3\beta_3$ ring. 
The remainder, $Q_\mathrm{motor}\equiv W-\Delta\mu$, dissipates irreversibly from the motor.

The motor molecule itself is not accessible by experiments.
Nevertheless, the heat dissipation from the nanosized motor $Q_\mathrm{motor}$ can be estimated from the thermodynamic quantities regarding the probe $W_\mathrm{ext}$ and $Q_\mathrm{probe}$ using the energy balance (\ref{eq:balance}). 
$W_\mathrm{ext}$ is easily calculated as noted, whereas it is usually difficult to measure heat $Q_\mathrm{probe}$ in a microscopic system subjected to thermal fluctuations. 
This is because, although heat can be defined as the microscopic energy exchange with the solvent \cite{Sekimoto2010}, the thermal fluctuating force is inaccessible using experimental manipulations. 
However, a nonequilibrium equality derived by Harada and Sasa \cite{Harada2005, Harada2006, Toyabe2007} enables us to calculate $Q_\mathrm{probe}$ from quantities that can be derived experimentally: the fluctuation and the response to a small external torque perturbation of the rotational rate (see Materials and Methods for details). 
Fluctuation and response are related by the fluctuation response relation (FRR) around the equilibrium state \cite{Kubo1991}. 
The FRR is generally violated far from equilibrium. 
Therefore, the extent of the FRR violation indicates how far the system is driven to nonequilibrium. 
The Harada and Sasa's relation relates the extent of the FRR violation to the heat dissipation at a nonequilibrium steady state.
This relation is valid in systems described by Langevin equations such as a colloidal particle in an aqueous solution.

In the following experiment, we evaluated $Q_\mathrm{probe}$ by Harada-Sasa relation, obtained $W$ from [\ref{eq:balance}], and compared it with $\Delta\mu$ to evaluate the energetics of mechanical-to-chemical free energy transduction. 
We also discuss the energetics of the chemical-to-mechanical transduction in the ATP-hydrolytic rotations.

\section*{Results}

\begin{figure}[!h]
\centerline{\includegraphics[scale=1]{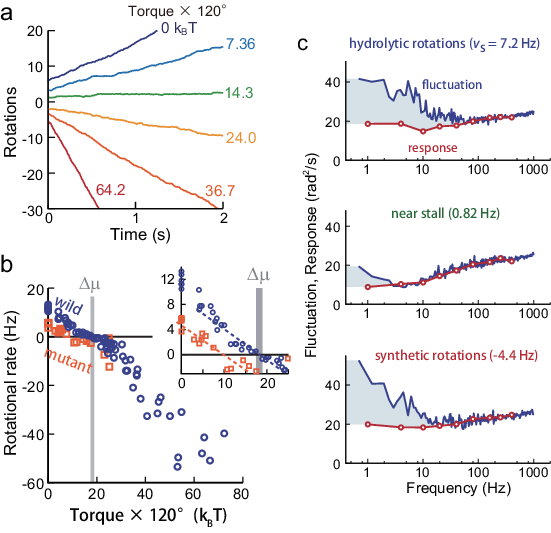}}
\caption{\label{fig:velocity}
{\bf a}, Rotational trajectory of the wild-type F$_1$-motor under an external load. 
{\bf b}, Rotational rate of the wild-type (circle) and mutant containing $\beta$G158A, T165S and Y341W\cite{Muneyuki2007Biophys} (square) in the presence of torque. 
The thick vertical line indicates $\Delta\mu=\Delta\mu^\circ+k_\mathrm{B}T\ln\mathrm{[ATP]/[ADP][P_i]}$ (17.5-18.9 $k_\mathrm{B} T$). 
The value of $\Delta\mu^\circ$ was calculated according to values adopted from the literature \cite{Guynn1973,Krab1992, Panke1997,Rosing1972,Toyabe2010PRL}.
The width of the line indicates the variation of $\Delta\mu$. 
The inset is the magnified view around the stalled state. 
The dashed lines were fitted by eye as guides. 
{\bf c}, Examples of fluctuations, $\tilde C(f)$, and responses, $2k_\mathrm{B}T\tilde R'(f)$. The torques multiplied by 120$^\circ$ are 4.15, 14.0, and 26.3 $k_\mathrm{B}T$, respectively, from top to bottom. 
The value of twice the shaded area corresponds to the integral in (\ref{eq:Harada-Sasa}).
}
\end{figure}

{\it Single-molecule response measurement and violation of the fluctuation-response relation.}
The $\gamma$-shaft's rotations were assessed using a large dimeric probe (diameter = 300 nm) at a relatively high ATP concentration (10 $\mu$M ATP, 10 $\mu$M ADP, 1 mM P$_\mathrm{i}$). 
Under this condition, the probe rotated smoothly without clear steps in the absence of external torque (Fig. \ref{fig:velocity}a). 
When we applied external torque to the probe using the electrorotation method in the direction opposite to the rotations, the rotational rate decreased (Fig. \ref{fig:velocity}a and b). 
At torques greater than $\Delta\mu/120^\circ$, the rotation was reversed, and the $\gamma$-shaft rotated in the ATP-synthetic direction. 
This is consistent with a previous result showing that the maximum work that the F$_1$-motor can exert per ATP cycle is similar to $\Delta\mu$ \cite{Toyabe2011}. 
Specifically, the thermodynamic efficiency of the F$_1$-motor is nearly 1 at the stalled state where the mean rotational rate vanishes.

The FRR is violated at low frequencies (Fig. \ref{fig:velocity}c) where the motor drives the probe to a nonequilibrium state \cite{Toyabe2010PRL}, whereas the FRR is held at high frequencies where the probe's motion is expected to behave as a freely rotating Brownian particle at equilibrium. 
The extent of the FRR violation, or twice that of the shaded area in Fig. \ref{fig:velocity}c, corresponds to the integral in Harada-Sasa relation (see Materials and Methods). 
When we increased the magnitude of the load from zero, the extent of the FRR violation decreased in the ATP-hydrolytic rotations, nearly vanished at the stalled state, and increased again in the ATP-synthetic rotations.

{\it Energetics of mechanical-to-chemical free energy transduction.}
In Fig. \ref{fig:Q}a, $Q_\mathrm{probe}$ evaluated by Harada-Sasa equality and $W_\mathrm{ext}\,(=\pm (\mathrm{torque})\times 120^\circ)$  are shown. 
$Q_\mathrm{probe}$ linearly decreased as we increased the load and vanished around the stalled state. 
This indicates that, at the stalled state, the probe behaved like a rotational Brownian motion at equilibrium in a 120$^\circ$-spacing periodic potential. 
In the ATP-synthetic rotations, the value of $Q_\mathrm{probe}$ increased as the torque increased further, mainly because the rotational rate increased with torque. 

In Fig. \ref{fig:Q}c, $Q_\mathrm{probe}$ is shown separately; the contribution from the steady motion $Q_\mathrm{s}$ and that from nonequilibrium fluctuations around the steady motion $Q_\mathrm{v}$, which correspond to the first and second terms of the rhs of the Harada and Sasa's relation (\ref{eq:Harada-Sasa}), respectively.
$Q_\mathrm{s}$ dominated $Q_\mathrm{v}$ in the present experimental condition.
An interesting observation is that $Q_\mathrm{v}$ as well as $Q_\mathrm{s}$ increases linearly as the torque increases.
The fraction of $Q_\mathrm{s}$ and $Q_\mathrm{v}$ are similar independent of the torque magnitude except around the stalling state.
The fraction of $Q_\mathrm{s}$ seems higher around the stalling state.
However, since $Q_\mathrm{s}$ and $Q_\mathrm{v}$ are small at this region, more precise measurement is necessary to test this dependency.

The difference $W=W_\mathrm{ext}-Q_\mathrm{probe}$ is the work received by the motor (Fig. \ref{fig:W}a). 
In the ATP-synthetic rotations, we found that $W=\Delta\mu$ over a broad range of torque, where $\Delta\mu$ is the thermodynamic limit for the work necessary to synthesize an ATP molecule. 
This implies that the motor's internal dissipation $Q_\mathrm{motor}=W-\Delta\mu$ is negligible and the F$_1$-motor can convert most of $W$ to $\Delta\mu$ even during rotations at a high rotational rate such as 50Hz. 
Note that the rate of the proton-driven rotations of the thermophilic F$_\mathrm{o}$F$_1$-ATP synthase is 3-4 Hz at saturated ADP and Pi concentrations at a room temperature \cite{Soga2011}.

When the motor rotates in the ATP-hydrolytic direction under weak external torque, $-W$ corresponds to the work performed by the motor transferred to the probe. 
A part of $-W$ is used to increase the potential of the probe against load $-W_\mathrm{ext}$, and the remainder, $Q_\mathrm{probe}$, dissipates. 
In Fig. \ref{fig:W}a, we show that $W$ is similar to $-\Delta\mu$ in good agreement with a previous result \cite{Toyabe2010PRL} that the motor can consume $\Delta\mu$ in the rotational degree of freedom with negligible internal irreversible heat production even during rotations. 
However, we observed a small deviation of $W$ from $-\Delta\mu$ at a small torque magnitude (Fig. \ref{fig:W}a inset), which was not observed in the previous experiment. 
$W$ increased with torque and reached $\Delta\mu$ around the stalled state.

We also measured $W$ of a mutant F$_1$-motor molecule with mutations around the nucleotide binding site ($\beta$T165S and $\beta$Y341W) and the $\beta$-subunit's hinge region ($\beta$G181A) \cite{Muneyuki2007Biophys} (Fig. \ref{fig:Q}b and \ref{fig:W}b). 
This mutant produced a smaller maximum work than the wild-type, supposedly caused by weak binding of ATP \cite{Toyabe2011}. 
Figure \ref{fig:W}b shows that $|W|$ is significantly less than $|\Delta\mu|$; the mutant produces a finite amount of internal dissipation.

\section*{Discussion}


Here we evaluated the energetics of the free-energy conversion by the F$_1$-motor during rotations far from a quasistatic process in mechanical-to-chemical and chemical-to-mechanical energy transduction. 
This was achieved by combining a single molecule response measurement mimicking the mechanical coupling of F$_\mathrm{o}$F$_1$-ATP synthase with a nonequilibrium theory. 
Our main finding is that, in the ATP-synthetic rotations, the internal dissipation from the motor is negligible even during rotations far from a quasistatic process. 
In the ATP-hydrolytic rotations, we observed a finite amount of the internal dissipation from the motor, which was not observed in the previous experiment. 

In case the probe and the $\gamma$-shaft are thermally insulated in that the heat transfer between them is negligible, $\Delta\mu/W\le 1$ in the ATP synthesis and $W/\Delta\mu\le 1$ in the ATP hydrolysis because $\Delta\mu$ is the thermodynamic limit for the work necessary to synthesize an ATP molecule and the thermodynamic limit for the work extractable from an ATP hydrolysis.
If this is the case, the present results, $\Delta\mu\simeq W$ in both the ATP synthesis and ATP hydrolysis, imply that the F$_1$-motor is a highly efficient motor working at almost 100\% efficiency even within finite-time.
The thermal insulation between the probe and the shaft is supposedly valid when the linker connecting the probe and the shaft is sufficiently flexible and $\Delta\mu$ and ATP concentration are not extremely small\cite{Zimmermann2012, Kawaguchi2014} although detailed simulations with explicit separation of the degrees of freedom of the motor and probe are necessary to deduce the exact conditions. 
Otherwise, the probe and the $\gamma$-shaft move in tandem to some extent and the heat dissipated from the system cannot be separated well into $Q_\mathrm{motor}$ and $Q_\mathrm{probe}$; for example, a part of the energy absorbed by the motor to overcome an energy barrier for a chemical reaction may dissipate through the probe's motion as a part of $Q_\mathrm{probe}$.  
$W$ evaluated by (\ref{eq:balance}) is no longer work that is transferred to the motor, and $\Delta\mu/W$ can be greater than 1 \cite{Kawaguchi2014, Wang2001EPL, Zimmermann2012}.
The spring constant of the linker is measured in ref\cite{Okuno2010} although their probe is different from this work.
We expect that future detailed simulation will reveal the implications of the intriguing balance $\Delta\mu\simeq W$ obtained in this work.

In the present study, we assumed tight coupling between ATP synthesis/hydrolysis and the 120$^\circ$ rotation and compared $W$ with $\Delta\mu$ to assess the efficiency. 
Although, the 120$^\circ$ rotational step and the ATP hydrolysis are tightly coupled \cite{Rondelez2005, Yasuda1998}, this has yet to be firmly established for ATP synthesis. 
The results of a pioneering experiment on the forced rotations of the $\gamma$ shaft that were induced using magnetic tweezers imply that less than one ATP molecule is synthesized in each 120$^\circ$ synthetic-direction rotation \cite{Rondelez2005}. 
In that study, the tweezers were rotated at a high rate of 10 Hz. 
Magnetic tweezers impose a trapping force, in contrast to the constant torque imposed by the electrorotation method. 
At a high manipulation rate the angular positions of the probe and tweezers can be apart. Then, magnetic tweezers can impose a very strong torque, which may induce slippage of the mechanochemical coupling. 
The results of a more recent experiment using magnetic tweezers manipulated at a low rate of 0.05-0.8 Hz suggests that attachment and detachment of nucleotides and the angular position of the $\gamma$ shaft are tightly coupled \cite{Adachi2012}. 
Further, the previous result showing that the maximum work of the F$_1$-motor is similar to $\Delta\mu$ over a broad range of $\Delta\mu$, suggests tight coupling \cite{Toyabe2011}. 
The mutant's finite internal dissipation implies that it fails to completely couple mechanical rotation and ATP hydrolysis and synthesis. 
Some externally-forced mechanical steps may not accompany ATP-synthetic reactions, and some ATP hydrolytic reactions may not accompany mechanical steps.

Macroscopic engines operated at a finite rate inevitably generates turbulence, and additional energy dissipates through microscopic degrees of freedom as irreversible heat. 
In contrast, the F$_1$-motor is itself microscopic and possibly utilizes thermal fluctuations. 
Therefore, some of the microscopic degrees of freedom might not be hidden and are accessible to the F$_1$-motor. 
In the ATP-hydrolytic rotations, previous studies suggest that the F$_1$-motor shifts the mechanical potentials discontinuously depending on the $\gamma$-shaft's angular position \cite{Toyabe2012, Watanabe2011}. 
Only by nanosized machines can perform such an operation to minimize irreversible heat within a finite-time. 
This highlights the remarkable property of nanosized engines. 
Further studies will be required to elucidate the molecular mechanism of the highly efficient mechanochemical couplings of ATP-synthetic rotations.

\section*{Materials and methods}

\begin{figure}[!h]
\centerline{\includegraphics[scale=1]{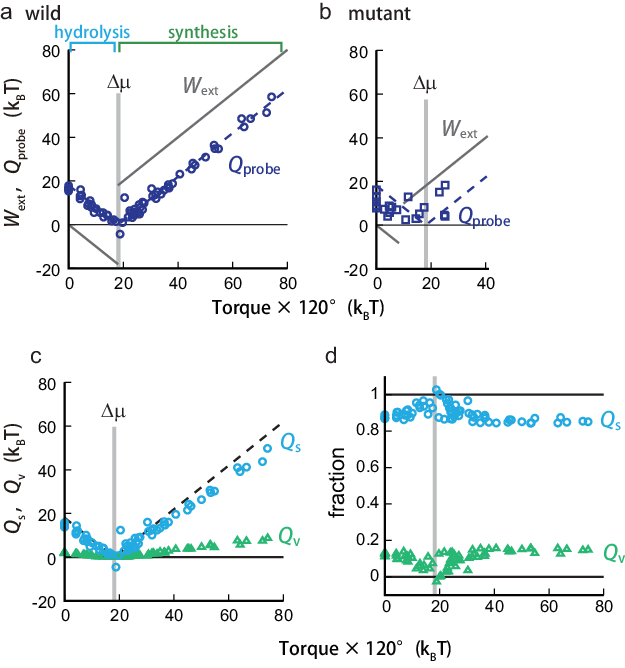}}
\caption{\label{fig:Q}
Heat dissipation through the probe. 
{\bf a}, {\bf b}, Amount of heat dissipation, $Q_\mathrm{probe}$ (circle and square) and the potential increase against an external load, $W_\mathrm{ext}$ (solid line), of wild-type ({\bf a}) and mutant ({\bf b}) F$_1$-motors. 
$W_\mathrm{ext}$ is calculated as the torque multiplied by 120$^\circ$ with the sign depending on the rotational direction, that is, positive in the ATP synthetic rotations and negative in the ATP hydrolytic rotations. 
69 points of 19 wild-type molecules and 20 points of 9 mutant molecules are shown.
We excluded one aberrant point due to the vanishingly small rotational rate from the graphs displayed in {\bf a} (torque $\times$ 120$^\circ$ and $Q_\mathrm{probe}$ are 20.0 $k_\mathrm{B}T$ and -40.5 $k_\mathrm{B}T$, respectively).
{\bf c}, Amount of heat dissipation of wild-type molecules was separately plotted as the heat contributed by the steady rotations $Q_\mathrm{s}$ and by the nonequlibrium fluctuations $Q_\mathrm{v}$, which correspond to the first and second terms of the rhs of the Harada-sasa equation (\ref{eq:Harada-Sasa}), respectively.
{\bf d}, Fraction of $Q_\mathrm{s}$ and $Q_\mathrm{v}$ of wild-type molecules.
}
\end{figure}

\begin{figure}[!h]
\centerline{\includegraphics[scale=1]{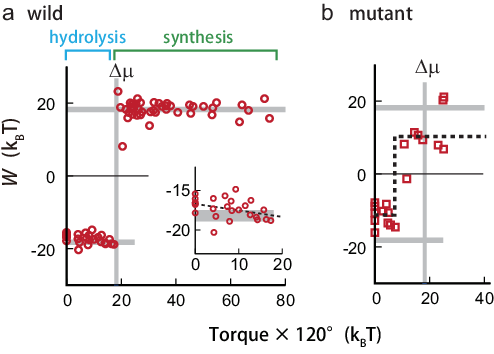}}
\caption{\label{fig:W}
Work performed on the motor for the wild type ({\bf a}) and mutant ({\bf b}). 
$W=W_\mathrm{ext}-Q_\mathrm{probe}$. 
The linear fits of the curves are $-16.7-0.082x$ in the hydrolytic rotations (dashed line in the inset of {\bf a}) and $18.2+0.00286x$ in the synthetic rotations (not shown), where $x$ is the torque multiplied by 120$^\circ$. 
The dashed line in b was fitted by eye. 
We excluded one aberrant point as noted in the caption of fig. \ref{fig:Q}.
}
\end{figure}

{\it Single molecule response measurement.}
 The experimental setup is essentially the same as that in the previous studies \cite{Toyabe2012, Toyabe2010PRL, Toyabe2011, Watanabe-Nakayama2008}. F1 molecules derived from a thermophilic Bacillus PS3 with mutations for the rotation assay (His$_6$-$\alpha$C193S/W463F, His$_{10}$-$\beta$, $\gamma$S107C/I210C, denoted as wild type) \cite{Rondelez2005} or a mutant with $\beta$T165S, $\beta$Y341W, and $\beta$G181A\cite{Muneyuki2007Biophys} were adhered on a cover slip functionalized by Ni$^{2+}$-NTA. Rotations of the $\gamma$ shaft were probed by streptavidin-coated dimeric polystyrene particles (diameter = 300 nm, Seradyn) attached to the biotinylated $\gamma$ shaft in a buffer containing 5 mM MOPS/KOH, 10 $\mu$M MgATP, 10 $\mu$M MgADP, 1 mM Pi, and 1 mM MgCl$_2$ (pH 6.9). Observation was performed on a phase-contrast upright microscope (Olympus) with a 100$\times$ objective and a high-speed camera (Basler) at 2,000 Hz. The data including a long pause presumably due to the MgADP-inhibited state are excluded from the analysis. We applied torque on the probe by using rotating electric field at 15 MHz generated with the quadrupolar electrodes patterned on the glass surface of the chamber \cite{Berry1995, Toyabe2011, Toyabe2010PRL, Toyabe2010b, Washizu1991}. The torque magnitude was controlled by controlling the electrodes voltage.
In Fig. \ref{fig:Q} and \ref{fig:W}, 69 points of 19 wild-type molecules and 20 points of 9 mutant molecules are shown.
 
{\it Fluctuation, response, and Harada-Sasa equality.}
Let $C(t)=\langle\left[v(t)-v_\mathrm{s}\right]\left[v(0)-v_\mathrm{s}\right]\rangle$ and $R(t)$ be the fluctuation and the response function against small external torque, respectively, of the probe's rotational rate $v(t)$ around the mean rotational rate $v_\mathrm{s}$. 
$R(t)$ is defined as $\langle v(t)-v_\mathrm{s} \rangle_N=\int^\infty_{-\infty} ds R(t-s)N(s)$, where $\langle\cdot\rangle_\mathrm{N}$ is the ensemble average under a sufficiently small probe torque $N(t)$. Because of the causality, $R(t)=0$ if $t<0$. Around an equilibrium state, $C(t)$ and $R(t)$ are related by the fluctuation response relation (FRR): $C(t)=k_\mathrm{B} TR(t)$ \cite{Kubo1991}. The FRR is generally violated far from equilibrium. 
The equality by Harada and Sasa relates the extent of the FRR violation to heat dissipation at a nonequilibrium steady state \cite{Harada2005, Harada2006, Toyabe2007}. 
In the frequency space, the equality is expressed as 
\begin{equation}\label{eq:Harada-Sasa}
Q_\mathrm{probe}=\frac\Gamma{3v_\mathrm{s}}\left[v_\mathrm{s}^2+\int^\infty_{-\infty} df[\tilde C(f)-2k_B T\tilde R'(f)] \right]\equiv Q_\mathrm{s} + Q_\mathrm{v},     
\end{equation}
where $\tilde C(f)$ and $\tilde R(f)$ are the Fourier transforms of $C(t)$ and $R(t)$, respectively, at frequency $f$, and $\tilde R'(f)$ is the real part of $\tilde R(f)$. 
$Q_\mathrm{s}\equiv \Gamma v_\mathrm{s}/3$ and $Q_\mathrm{v}\equiv \Gamma/(3v_\mathrm{s})\int^\infty_{-\infty} df[\tilde C(f)-2k_B T\tilde R'(f)]$ correspond to the dissipation through steady rotations and that through nonequilibrium fluctuations in a 120$^\circ$ rotation.
This relation is valid in the systems described by Markovian Langevin equations.
The FRR becomes $\tilde C(f)=2k_\mathrm{B}T\tilde R'(f)$ in the frequency space, $\Gamma$ is the rotational frictional coefficient, and 3$v_\mathrm{s}$ is the mean stepping rate that corresponds to the rate of ATP hydrolysis or synthesis. 
$\tilde C(f)$ was calculated from the rotational trajectories by a fast Fourier transform method and Wiener-Khintchine theorem. 
$\Gamma$ was obtained by taking the average of $\tilde C(f)$ around 300 Hz since the FRR, $\lim_{f\to\infty}\tilde C(f)=2k_\mathrm{B}T/\Gamma$, is supposed to hold in such a high frequency region \cite{Toyabe2011, Toyabe2010PRL}. 
$\Gamma$ was 0.075$\pm$0.012 $k_\mathrm{B}T\,\mathrm{s/rad^2}$ (mean$\pm$SD, N=70).
For evaluating $\tilde R(f)$, we added a small torque $N(t)=N_0\sum_\mathrm{i}\sin(2\pi f_\mathrm{i} t)$ , where $f_\mathrm{i}$ = 1, 4, 10, 20, 40, 80, 160, 250, and 400 Hz, in addition to the constant load. 
$N_0$ is unknown a priori. 
We measured $\langle v(t)\rangle_\mathrm{N}$, performed a Fourier transform, and obtained $R(f_\mathrm{i})$ $N_0$ at multiple frequencies $f_\mathrm{i}$. Then, we obtained $N_0$ by comparing $\tilde C(f)$ and $\tilde R(f) N_0$ around at high frequency regions because of the FRR \cite{Toyabe2011, Toyabe2010PRL}. The frequency regions for calibration were determined by eye (around 300Hz typically). Finally, we obtained $R(f_\mathrm{i})$. 
$N_0$ was 0.94$\pm$0.19 $\mathrm{k_B T/rad}$ (mean$\pm$SD, N=70). We calculated the integration in (\ref{eq:Harada-Sasa}) between -400 Hz and 400 Hz. The contribution from the outside of this region is supposed to be negligible because of the FRR at high frequencies. We omitted data without apparent convergence of $\tilde C(f)$ and $\tilde R'(f)$ at the high frequency region (8 out of 78). 
When torque multiplied by 120$^\circ$ was greater than 80 $k_\mathrm{B}T$, the FRR was violated at high frequencies even around 300 Hz, where the torque magnitude was calibrated according to the FRR.

\begin{acknowledgments}
We thank valuable discussions with Kyogo Kawaguchi, Takahiro Sagawa, Masaki Sano, Shin-ichi Sasa, and Hiroshi Ueno. 
This work is supported by a Grant-in-Aid for Scientific Research on Innovative Areas "Fluctuation \& Structure" (No. 26103528) from the Ministry of Education, Culture, Sports, Science, and Technology of Japan. 
\end{acknowledgments}

\bibliographystyle{iopart-num}
\bibliography{njp}


\newpage

\end{document}